\documentclass[prl,aps,amssymb,amsfonts,amsmath,showpacs,twocolumn]{revtex4-1}
\pdfoutput=1

\usepackage{textcomp}
\usepackage{graphicx}

\usepackage{color} 

\definecolor{darkblue}{rgb}{0,0,0.5}
\definecolor{lila}{rgb}{0.3,0,0.3}
\definecolor{turq}{rgb}{0,0.1,0.4}
\definecolor{lightblue}{rgb}{0.7,0.7,0.9}

\usepackage{url} 

\usepackage[pdftex,
  colorlinks=true,
  backref=page,
  linkcolor=darkblue, 
  filecolor=red,
  citecolor=turq, 
  urlcolor=lila, 
  pdftitle={Detection of Single Molecules Illuminated by a Light-Emitting Diode},
  pdfauthor={Ilja Gerhardt, Lijian Mai, Antia Lamas-Linares, Christian Kurtsiefer},
  pdfsubject={},
  pdfkeywords={single molecules; fluorescence microscopy; light-emitting diode; LED; signal to noise ratio; single photon detection},
  pdfpagelabels=true, 
  breaklinks=true,
  plainpages=false,
  backref=false,
  bookmarks,
  bookmarksnumbered=true]{hyperref}

\begin{document}
\title{Detection of Single Molecules Illuminated by a Light-Emitting Diode}

\author{Ilja Gerhardt}\email{ilja@quantumlah.org}\altaffiliation{present address: Low Temperature Group, Chemistry Department, University of British Columbia, 2036 Main Mall, Vancouver, B.C. Canada V6T 1Z1}
\author{Lijian Mai}
\author{Ant\'ia Lamas-Linares}
\author{Christian Kurtsiefer}
\affiliation{CQT, Centre of Quantum Technologies, 3 Science Drive 2, 117543 Singapore}

\begin{abstract}
Optical detection and spectroscopy of single molecules has 
become an indispensable tool in biological imaging and sensing. Its
success is based on fluorescence of organic dye molecules under 
carefully engineered laser illumination. In this paper we
demonstrate optical detection of single molecules on a wide-field 
microscope with an illumination based on a commercially available,
green light-emitting diode. The results are directly compared with
laser illumination in the same experimental configuration. The setup
and the limiting factors, such as light transfer to the sample,
spectral filtering and the resulting signal-to-noise ratio are
discussed. A theoretical and an experimental approach to estimate
these parameters are presented. The results can be adapted to other
single emitter and illumination schemes.
\end{abstract}


\maketitle


\section{Introduction}

The invention of the laser in the 1960s was a key evolution in the
path towards optical single molecule detection. Already in 1976
\textsc{Hirschfeld} performed an experiment, which was an important
step towards this goal~\cite{hirschfeld01}. Using a laser to excite
a fluorescently doped sample and detecting the spectrally filtered
light on a photomultiplier, he was able to see the fluorescent
fingerprint of a cluster of molecules. In the 1980s \textsc{Moerner}
and \textsc{Kador} succeeded in the first optical detection of
single pentacene molecules~\cite{moerner01}, but this new technique
only became important for biology and sensing in the 1990s when the
experiments were extended to work at room temperature. These
experiments rely on efficient discrimination of the excitation laser
light from the molecule's red shifted fluorescence~\cite{orrit02}.
Since then, single molecule spectroscopy has become a valuable tool
to overcome ensemble averaging over many emitters and to perform
microscopy at a sub-diffraction limited
scale~\cite{betzig07,Hell2003}. In terms of sensitivity the
detection of a single molecule of a certain compound represents the
ultimate limit.

The above mentioned experiments were performed using laser
illumination. The narrow linewidth and the coherent nature of the
laser emission allow for spectral discrimination and easy focussing.
Experiments performed using other light sources for single molecule
research are
rare~\cite{unger_biotechniques_1999,kuo,hattori_cl_2009}. In other
fields, e.g.,~in white light imaging or fluorescence microscopy, the
usage of non-laser sources is well established. At present, one of
the most interesting light sources for microscopy is the
light-emitting diode (LED), which underwent significant engineering
efforts in the last two decades. High-power LEDs are commercially
available, and could be thought of as compact and inexpensive
sources for single molecule detection and sensing applications, or
even for single photon generation. Their availability across the
spectral range from 280 to 1300~nm, their stable output, their
electrical insensitivity, and long lifetime make them attractive
alternatives to laser diodes for some applications.

In this paper we discuss the use of a commercially available LED as
an excitation source for single molecule studies. Unlike presented
before~\cite{kuo,hattori_cl_2009}, we extend the experiments to the
green part of the visible spectrum. The minimum irradiance to excite
and detect single molecules is estimated by comparing the expected
illumination levels to the nominal sensitivity of a camera and
single photon detectors.

A rigorous proof that a single molecule has been observed is only
possible by detecting the characteristic anti-bunched photon
statistics~\cite{basche01}. Prospects for the generation and
detection of single photons based on LED illumination are discussed.
Experimentally, we compare LEDs and laser excitation schemes side by
side. It is shown that single molecules can be imaged with LED
illumination and the influencing factors are presented. This extends
the work of \textsc{Kuo} and coworkers, which mention present
experimental findings on single molecule detection in the blue part
of the spectrum~\cite{kuo}.


\section{Material and Methods}

\begin{figure}[b]
\centering
\includegraphics[width=8.5cm]{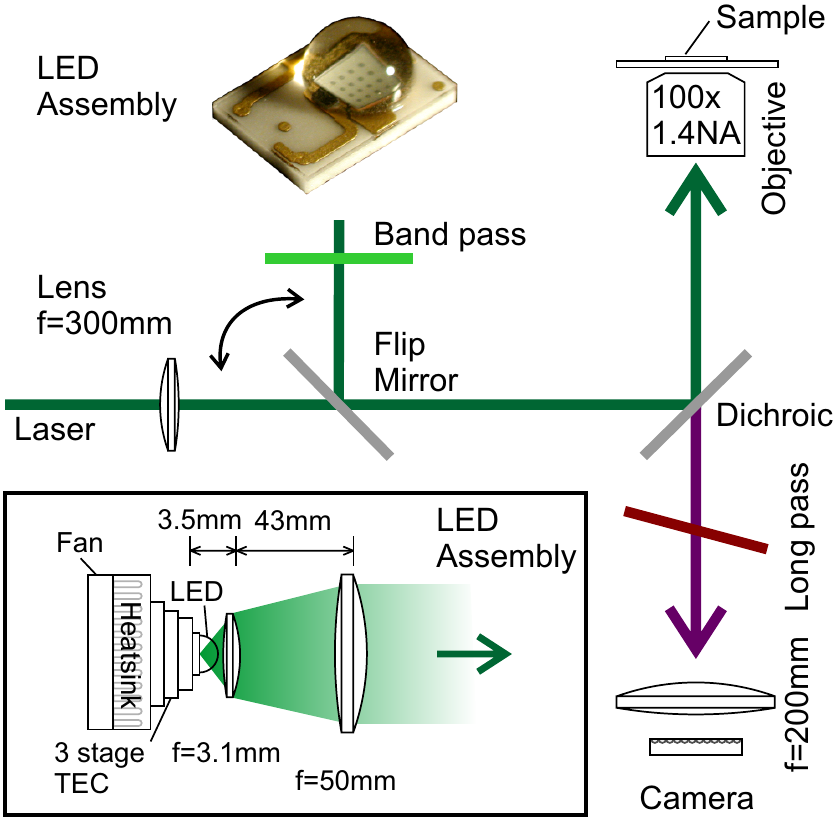}
\caption{Experimental Setup, consisting of a confocal microscope
(detection not shown) with wide-field configuration. A flip mirror
allows to switch between laser and LED type illumination. Inset: Two
lens LED-assembly, the LED is mounted with thermal grease directly
onto a 3 stage thermo-electric cooler (TEC), which is attached to a
fan-cooled heat sink.\label{fig:setup}}
\end{figure}

The sample for all further experiments and estimations consists of a
doped thin crystalline film of \emph{p}-terphenyl (\texttt{Aldrich})
which is spin coated on a microscopy coverslip~\cite{pfab}. As a
dopant, the well characterized fluorescent molecule terrylene was
chosen~\cite{kummer,kulzer_chemphys_1999}. The concentration of
terrylene molecules (\texttt{PAH Research}) is in the order of
$\approx$10$^{-10}$ to allow spatial separation. In an area of \linebreak 
10 $\times$ 10~$\mu$m$^2$ 1--20 molecules were observed. The
coverslips were cleaned by organic solvents in an ultrasonic bath
and afterwards transferred to a solution of 1 part sulfuric acid and
3 parts hydrogen peroxide (pira\~na solution) to clean organic
residues. The coverslips continue to be submerged in the solution
for storage until use and rinsed under water prior to spin coating
sample preparation.

All our single molecule studies were performed on a custom-made
inverted microscope which can be configured for wide field or
confocal imaging, and can use LED or laser illumination in either
configuration (Figure~\ref{fig:setup}). As a well characterized
light source a frequency doubled Nd:YAG laser (532~nm) was coupled
into the excitation path via a single mode optical fiber. To allow
microscopy in wide-field mode, a f = 300~mm lens was placed in the
optical path to focus the light into the backfocal plane of the
microscope objective and to produce a Gaussian shaped illuminated
area on the sample (full width, half maximum, FWHM $\approx$ 8~$\mu$m). 
Illumination and detection were performed through a
100$\times$, 1.4 NA objective (\texttt{Olympus}, \texttt{UPlanSApo}).
The detection in wide-field mode was realized by a low-cost
commercial astronomical camera (\texttt{Watec}, \texttt{Wat-120N+},
CCD: \texttt{SONY} \texttt{ICX-419ALL}). The image was captured with
a 200~mm camera objective (\texttt{AF} \texttt{Nikkor}). A single pixel on
the camera corresponds to the width of a standard deviation of the
diffraction limited spot ($\sigma \approx$120~nm) on the sample,
assuming the diffraction limited spot to show a FWHM of
$\approx$300~nm. Thereby the pixel to spot size ratio is 1, as
defined in~\cite{thompson_biophysjourn_2002}, which is optimal for
localizing single molecules. To record images, a video grabber card
with a digitizing resolution of nominal 8~bit was used. The
integration time of the camera could be set to a maximum value of
10.24~s.

For experiments using an LED, a flip mirror was introduced into the
illumination path, such that either the laser or the LED light was
passed to the sample, while the detection path remained unchanged.
This allows for a comparison of the LED-based results with the well
characterized laser illumination results. For laser illumination the
detection was performed in confocal and wide-field configuration,
whereas for LED illumination only wide-field images were recorded.

\begin{figure}[t]
\centering
\includegraphics[width=8.5cm]{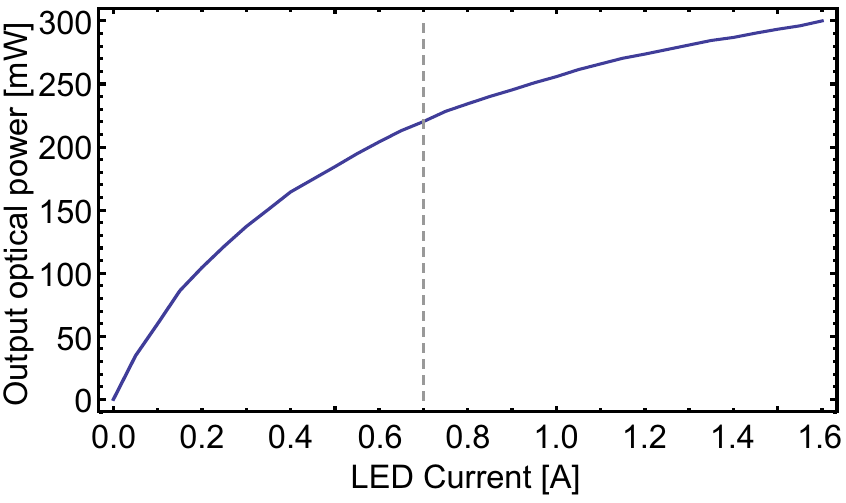}
\caption{Measured light emission directly in front of light-emitting
diode mounted in the diode assembly. The dashed line shows the
nominal maximal current of 700~mA.\label{fig:powercurrent}}
\end{figure}

\begin{figure}[t]
\centering
\includegraphics[width=9cm]{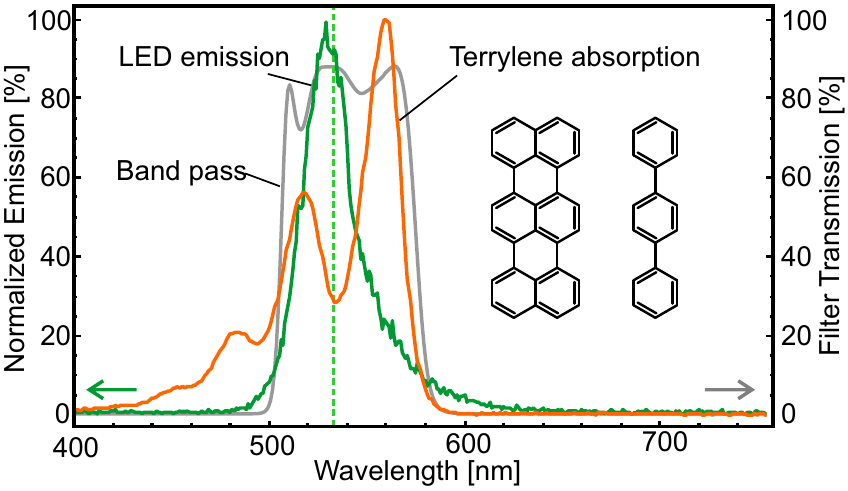}
\caption{Absorption spectrum of terrylene and the emission spectrum
of the unfiltered LED. The dashed line represents the wavelength of
the frequency-doubled Nd:YAG laser. The irradiance of one to the
other shows a by 40\% larger value for the LED illumination. The
larger spectral overlap allows a more efficient excitation. Inset:
Terrylene (left) and the matrix molecule \emph{p}-terphenyl
(right).\label{fig:spectra}}
\end{figure}

The LED illumination source selection was based on the spectral
overlap with terrylene absorption. Therefore commercially available
green LEDs with a center wavelength around 535~nm were evaluated.
The one with the highest irradiance was chosen for further
experiments (\texttt{Lumin\-leds} \texttt{Luxeon} \texttt{Rebel}, \texttt{LXML-PM01-0080}, InGaN). From a die
surface of 1.6 $\times$ 1.6~mm$^2$ an optical power of 240~mW at the
nominal maximum current of 700~mA was detected on an optical power
meter placed directly in front of the diode. The current could be
increased to 1.6~A, resulting in P$_{\rm out}$ = 300~mW
($\approx$180~lm, Figure~\ref{fig:powercurrent}), with proper
cooling and by sacrificing the device lifetime. The light yield in
this high power range is about 5\%. On the die surface the emitted
power corresponds to an exitance of 120~kW/m$^2$. The wavelength
shift over the entire current range was less than 5~nm from the peak
wavelength of \linebreak $\lambda$ = 530~nm (see
Figure~\ref{fig:spectra}). The spectral radiant exitance
$M_{\lambda=530nm}$ is 3000~W/(m$^2$~nm).

The thermal management of high power LEDs plays an important role
for the device lifetime. Unlike in
Reference~\cite{kuo,hattori_cl_2009} we utilize LEDs in the green
region in the spectrum. Green LEDs have the lowest light yield of
LEDs in the visible spectrum, whereas the earlier described
\cite{kuo,hattori_cl_2009} blue/UV LEDs have efficiencies up to
20\%~\cite{peter_pss_2009}. Our presented experiments require a much
more careful designed cooling of the LED, but also might be extended
to dyes which are presently more relevant in biological imaging,
such as Cy3, Cy5 and other Alexa-Dyes in the red region of the
spectrum. To allow an extended power supply to the LED, a three
stage thermo-electric cooler (\texttt{Ferrotec},
\texttt{9530/119/045 B}) was used to cool the LED base to
temperatures below 0~\textdegree C. Under operating conditions,
condensation was inhibited due to a higher thermal load of the LED.

Spectral filtering in the excitation path was performed with a
500--580~nm band pass filter, attached directly to the LED assembly
(\texttt{Photonik}, \texttt{Singapore}, Figure~\ref{fig:spectra}). This
filter was needed to suppress higher wavelength components of the
LED emission, which were leaking through the detection filter. The
integral LED transmission through the band pass filter was measured
to be 70\%.

\begin{figure}[b]
\centering
\includegraphics[width=8.5cm]{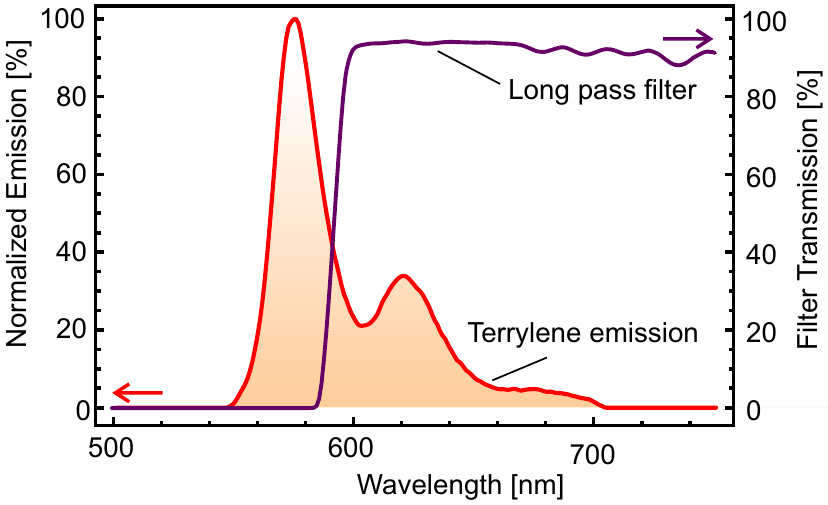}
\caption{Emission spectrum of terrylene molecules. The optimal
spectral filtering utilizes a similar slope to the emission spectrum
of the molecule. In our experimental configuration a long-pass
filter was slanted to match the excitation filter with a falling
slope around 585~nm.\label{fig:emission}}
\end{figure}

The detection path was equipped with a dichroic (50\%
Transmission/Reflection at 567~nm, \texttt{Thorlabs} \texttt{DMLP}
\texttt{567}) and a long pass filter with a nominal cut-off
wavelength of 640~nm (\texttt{Omega} \texttt{Optical} \texttt{640AELP}). The latter
was slanted to match the cut-off wavelength of the exciter (see
Figure~\ref{fig:emission} for the transmission of the slanted
filter). The angle was tuned by observing the camera images for
minimal background luminescence. The use of a dichroic mirror also
reduces the transmission of excitation light towards the detector; a
complementary pair of long pass and short pass filters should be
sufficient, with the long pass filter at the detector and the short
pass filter at the light source.

An optimal spectral filtering scheme is important to distinguish
between excitation light and detected fluorescence, thus maximizing
the signal to background ratio. Ideally, the filters would transfer
all excitation light to the sample and simultaneously pass the
entire resulting fluorescence to the detector. Simultaneously these
filters have to block the excitation light scattered off the sample
into the detection path.  If a short pass/long pass filter set has
complementary step transmission spectra around a cutoff wavelength
$\lambda_{\rm cut}$, and the suppression of backscattered excitation
light is taken care of, one can optimize the signal from a molecule
by varying the cutoff wavelength for broadband sources like LEDs.
For that, we combine the spectral power density of the source
$l(\lambda)$, the transmission $f_{\rm exc}(\lambda)$
through all excitation filters, and the normalized absorption
spectrum $a(\lambda)$ of the molecule to an effective excitation
flux $\phi_{\rm excitation}$ by integration over all excitation
wavelengths $\lambda_e$:

\begin{equation}
\phi_{\rm excitation}=\int_{\lambda_{\rm min}}^{\lambda_{\rm cut}}
l(\lambda_e) f_{\rm exc}(\lambda_e) a(\lambda_e) \; d\lambda_e
\;\label{eqn:filterlinetop}
\end{equation}
It turns out that the normalized excitation flux for our LED is
about 40\% larger than for the laser at 532~nm due to a better
spectral match.

If we assume that the fluorescence spectrum is independent of its
excitation spectrum, which is the case for simple optical
fluorescence configuration, the detected power is proportional to
the product of this excitation flux. A detection path response
$\phi_{\rm detection}$ combines the emission spectrum $e(\lambda)$
of the molecule, normalized to $e=1$ in its maximum, the 
transmission $f_{\rm det}(\lambda)$ through all detection filters,
and the detector response $r(\lambda)$ for all detection wavelengths
$\lambda_d$:
\begin{equation}
\phi_{\rm detection}=\int_{\lambda_{\rm cut}}^{\lambda_{\rm max}}
\;e(\lambda_d) f_{\rm det}(\lambda_d) r(\lambda_d) \; d\lambda_d
\label{eqn:filterline}
\end{equation}
For the combination of the LED and Terrylene emission/absorption
spectra in our experiment (see Figures~\ref{fig:spectra} and
\ref{fig:emission}), a value of $\lambda_{\rm cut}=565~\mbox{nm}$
maximizes the product $\phi_{\rm excitation}\cdot\phi_{\rm
detection}$ and hence the detection signal.

One of the main advantages of coherent illumination is the
constructive interference at the (laser) focus, reaching a maximum
light intensity to effectively excite the molecule. Laser light can
be focussed to a size approximately half the wavelength
($\lambda/2$), which is not possible for LED light. The limited
phase-space density of incoherent light results in a larger
illumination area and correspondingly lower irradiance. Usual LEDs
exhibit a \textsc{Lambertian} spatial emission profile and are
equipped with a solid immersion medium, consisting of either
polycarbonate, acrylic glass, or, as in the used LED, silicon rubber
for high power devices. This effectively increases light emission
out of the die and concentrates light in the forward direction.
Depending on the manufacturing accuracy of the die and the immersion
medium the emission profile can vary significantly. To adapt the
emission characteristics of the LED to the imaging geometry of the
microscope objective, further collimation was necessary. For our
experiments a high NA outcoupling lens (Geltech, C330TME-A,
f = 3.1~mm, 0.68 NA) is placed very close to the LED. For this
lens/LED combinations the solid immersion assembly was limiting the
minimum focussing distance, thus limiting the effective outcoupling
NA. The calculated radiance in one hemisphere is 20~kW/m$^2$sr,
taking the power of 300~mW, originating from an area of 1.6 $\times$
1.6~mm$^2$. With the aspheric lens in place and an assumed effective
NA of 0.5, we expect a high outcoupling efficiency of 80\%, if the
emitted profile is purely \textsc{Lambertian}. The measured power
behind the aspheric lens is 25\% of the maximal measured power of
the system right behind the die surface of the LED, suggesting a
deviation of the \textsc{Lambertian} intensity profile and an
effective lower NA. With these values, we calculate a radiance of
30~kW/m$^2$sr. Assuming a focussing angle onto the sample of 3.9~sr
(1.4 NA), this corresponds an maximum irradiance of 120~kW/m$^2$ on
the sample surface. The outcoupling aspheric lens was used in
combination with a 50~mm achromatic lens to channel emitted light
into the forward direction. The measured equivalent focal length of
the system was 10~mm.

Due to the finite extension of the LED die and the incoherent light
emission, the beam cannot be fully collimated. Initial attempts to
filter the modes utilizing a single mode fiber were resulting in an
outgoing power in the nW range and were not further pursued, such
that wide-field illumination had to be used. To ensure an optimal
transfer towards the sample and reduce clipping, the LED assembly
was placed in close proximity to the microscopy setup. Initially the
die was imaged to the backfocal plane of the microscope objective to
reduce clipping in the optical path. This configuration resulted in
a blurry die imaged on the sample and showed a lower intensity at
the focal plane as independently measured with fluorescent beads.
For later experiments the die was focused onto the coverslip such
that its structure was visible on the wide-field camera if the
spectral filters were removed.

\section{Results and Discussions}
\subsection{Estimation of Minimum Irradiance for Detecting Single Molecules}

As a first approach to estimate the minimum irradiance needed to
detect single molecules which are illuminated with an LED we
introduce a transfer expression. This relates the incident light to
the detected outcome and compares it to the sensitivity and noise
levels in the system. The interaction of light with a single emitter
is determined by the effective absorption cross section of the
emitter and the irradiance on this area, {\em i.e.},~the effective
field strength at the location of the emitter and the excitation
probability. The extinction cross section of a single molecule is in
the order of $\sigma_{\rm abs} \approx1 \times
10^{-15}$~cm$^2$~\cite{kukura-2008,kulzer_chemphys_1999}. By tightly
focussing light down to a diffraction limited spot, diameter
$\approx$300~nm, we have an effective overlap $\sigma_{\rm eff}$ of
$1.5 \times 10^{-6}$, {\em i.e.},~only one photon of 10$^6$ is
exciting the molecule, whereas the remaining light is non
interacting or has an option to interact with the environment,
leading to unwanted background. Depending on the fluorescence
quantum yield $\Phi_{\rm fl}$, and the molecules branching ratio
$\alpha_{\rm branch}$, only a fraction of absorbed photons leads to
a red shifted emission which is later detectable. The integrated
detection efficiency $\eta_{\rm det}$ for usual confocal microscopes
is usually estimated to be 1--5\%, and can be described by the
following terms: The geometrical pickup $\eta_{\rm geo}$ is mainly
determined by the numerical aperture of the microscope objective. In
our experimental configuration with a 1.4~NA objective, we cover a
half angle of 67\textdegree, corresponding to 30\% of the entire
emission in 4$\pi$. We detect light in the range of 590--700~nm,
such that 45\% of the spectral emission is transferred to the
detector ($\eta_{\rm spec}$). All filters in the detection path were
measured to have an integral transmission of better than 85\%
($\eta_{\rm filt}$). Finally the quantum efficiency of our detector
$\eta_{\rm qe}$ is in our detection range (590--700~nm) between 55\%
and 35\%, whereas most of the light is emitted at longer
wavelengths. This results in an effective detector quantum
efficiency of 50\%.

Assuming isotropic emission, the detected power $P_{\rm det}$ from
the molecule reads as
\begin{eqnarray}
P_{\rm det}&=&P_{\rm in} \times \sigma_{\rm eff} \times \Phi_{\rm fl} \times \alpha_{\rm branch} \times \eta_{\rm det}
\end{eqnarray}
with
\begin{eqnarray}
\eta_{\rm det}&=& \eta_{\rm geo} \times \eta_{\rm spec} \times
\eta_{\rm filt} \times \eta_{\rm qe}
\end{eqnarray}

\noindent
We assume the quantum yield of the molecule $\Phi_{\rm fl}$ to be
0.7~\cite{moerner04} and a detection efficiency $\eta_{\rm det}$ of
5\%. The branching ratio between red-shifted and resonantly
scattered photons is assumed to be 1, because we excite into a
higher vibrational level and consider the entire emission from the
first electronic excited state. It follows that an incident rate of
$1.5 \times 10^{10}$ photons/s (corresponding to 80~kW/m$^{2}$) onto
the absorption cross section is needed to observe a flux of 800
detectable photons/s on the CCD camera. We now relate this rate to
the detection sensitivity of the camera:

The camera's minimal illumination is $2 \times 10^{-5}$~lx, which
corresponds to $\approx$40 photons per pixel and second in a
spectral detection range of 590--700~nm. The measured
noise-equivalent power with 10.24~s integration time was 36 photons
per pixel. The molecule emits 15,000 photons per second, from which
800 per second are detectable, originating from a diffraction
limited spot, which is imaged such that an average of 20\% of its
detectable emission is imaged on a single pixel. This delivers an
illuminance of $\approx8 \times 10^{-5}$~lx or 160 photons per pixel
and second, four times above the nominal minimum camera sensitivity.

The figure of merit in all experiments is the signal-to-noise ratio,
which allows to differentiate between the actual signal and the
background noise. The intrinsic noise sources such as photon shot
noise, darkcount noise and background noise have to be included.

At the same irradiance levels as mentioned above we have a shot
noise level of $\sqrt{800}$~photons/s. This value can be directly
compared to the noise level of a single photon detector, such as an
avalanche photo diode (APD). The given minimum illumination of the
camera has to include the intrinsic signal to noise ratio and is
already covered by our sensitivity assumptions above. When blocking
the incident light on the camera and still ensuring the linearity
over the whole detection range, we detect a background level of
13\%, which would correspond to 270 dark counts per second on a
single photon detector. The measured RMS noise of 2\% corresponds to
36 photons per second. This value is approximately twice above the
shot noise limited value of $\sqrt{270}$ photons. Interestingly the
background level with LED illumination is 3--4 times higher than the
camera background, due to imperfections of the spectral filtering.
Since we are able to alternate between the two illumination options,
we are able to exclude this effect to result solely from the
background fluorescence of the sample. The intrinsic noise level of
the camera does not allow us to associate an increase of the noise
level due to the increased background and does not change
significantly within the detection range.

\subsection{Measurement of the Minimum Irradiance for Single Molecule Detection}

To experimentally determine the minimum irradiance levels with LED
excitation, we performed laser wide-field imaging and determined the
minimum amount of irradiance, needed to observe single molecules on
the camera. In these experiments, the integration time of the camera
(10.24~s) and the gain were increased to the maximum and the laser
power was subsequently reduced. At a laser irradiance threshold of
80~kW/m$^2$ single molecules were still observable on a camera
without any image postprocessing. The irradiance of the LED has
nominally the same value, and given the 40\% better spectral overlap
this value should be sufficient for imaging.

\begin{figure}[t]
\centering
\includegraphics[width=8.5cm]{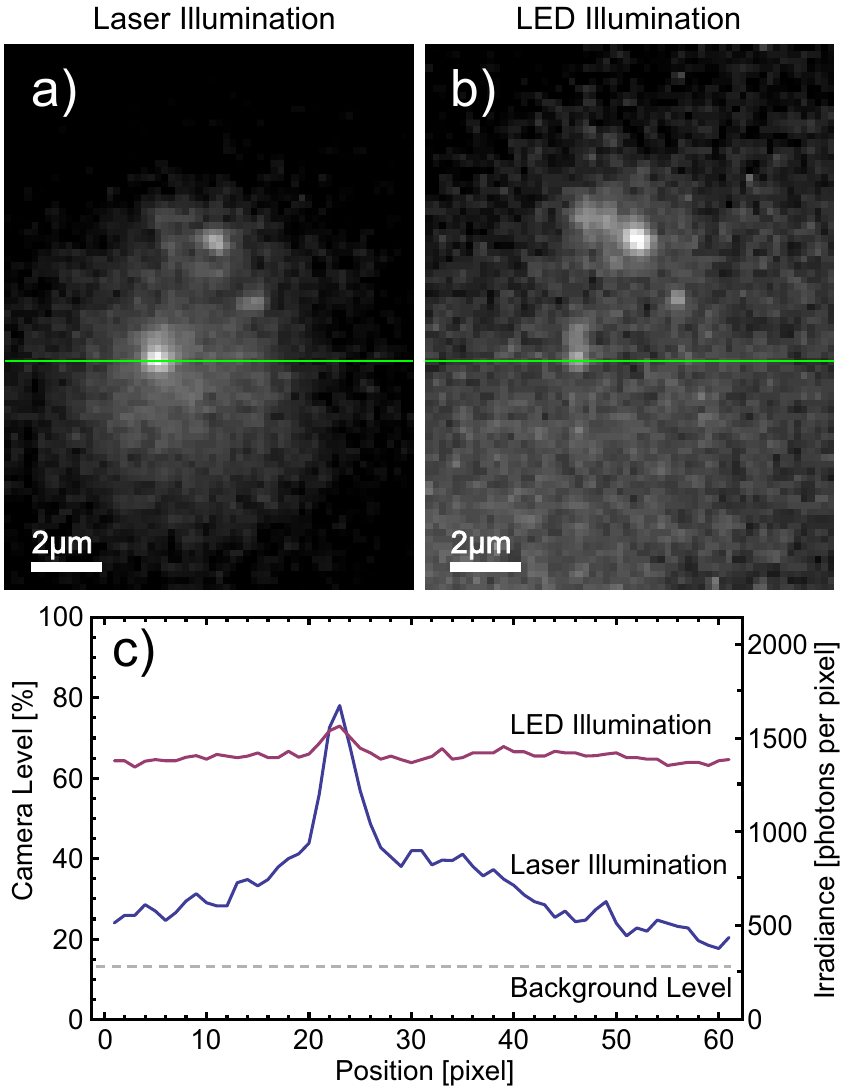}
\caption{Direct comparison of two wide-field images acquired by
laser (a) and LED illumination (b). The gray levels of the images
have been adapted to account for the increased background with the
LED illumination. Still the slightly weaker signal to noise ratio is
obvious. To have a direct comparison the camera levels are presented
uncorrected in figure (c). The background level of the camera is at
13\% (blocking the excitation light) and should be subtracted for
both illuminations. Light leakage through the filters increases the
LED illumination background level to more than
60\%.\label{fig:led_laser}}
\end{figure}

Before observations are carried out with LED illumination, the
spatial locations of individual molecules are determined using
wide-field laser illumination with sufficiently high irradiance. At
irradiance levels of about 80~kW/m$^2$ fluorescence blinking cannot
be observed due to the long integration time, whereas single step
bleaching from frame to frame strongly indicates the detection of
single molecule signals. After this, the excitation path was changed
and alternating images were recorded with laser and LED
illumination. An example is presented in
Figure~\ref{fig:led_laser}(a,b). The images are not processed, but
only a fraction of the dynamic range is shown. We determined the
laser wide-field illumination to be on a Gaussian area with FWHM of
$\approx$8~$\mu$m diameter. In the center the irradiance is about
80~kW/m$^2$, and molecules are still observable at lower intensities
in these images utilizing a smaller dynamical range. The effective
pixel size is increased by the frame grabber card to $\approx$200~nm
per pixel. In the measured microscope performance we achieve a
slightly larger patterns as expected for diffraction limited
microscopy. For LED illumination we achieve a FWHM of 330~nm,
whereas the laser illumination leads to a FWHM of 430~nm. We
attribute this deviation to a mechanical drift.

In Figure~\ref{fig:led_laser}(c) no corrections have been made and
the full camera range is shown. For the laser illumination we
observe the wide-field illuminated area and a flat illumination
background for LED illumination. The Gaussian intensity envelope in
the laser excitation corresponds to the wide-field spot generated by
the lens in the incident beam path.

For the laser and the LED illumination, we noticed a significant
background contribution originating from residual fluorescence of
the \emph{p}-terphenyl. Here we underline that the direct comparison
of the two illumination methods allows for a judgment on the sample
quality and to differ this from spectral leaking through the
excitation filters. In our experimental findings both parameters are
not negligible and lead to a two times higher background intensity
(see Figure~\ref{fig:led_laser}(c)).

\subsection{Detection of Single Molecule Signals}

Although being at the lower limit of the excitation intensity we
succeeded to detect single molecules with LED illumination in
wide-field mode with a signal to noise ratio of 3.5. The
corresponding signal to noise ratio generated by laser illumination
was 19. The signal to background ratio is 0.12 and 1, respectively.
To determine the presence of an object, the \textsc{Rose}'s
criterion in imaging suggests a signal to noise ratio above
5~\cite{Bushberg2001}, which is not given in our configuration with
LED illumination. However, with additional knowledge about the
emitter, namely its nanometer size, resulting in a diffraction
limited spot, we are able to determine single molecules against the
noise floor of our detection scheme. As usual, wide-field detection
allows us to use a localization algorithm to spatially localize the
single emitters to a sub-diffraction accuracy. For samples with
higher doping of fluorescent molecules the \textsc{Rose}'s criterion
would indeed limit the probability to determine single molecule
signals.

\section{Outlook}

The dual widefield/confocal configuration of this instrument is
optimized for quick localization of single emitters and subsequent
confocal imaging and measurement of individual molecules. We have
shown that the current setup is able to reliably identify single
molecules with both laser and LED illumination sources. To
rigorously prove the presence of a single emitter one has to detect
the single photon emission. Therefore the photon-photon
autocorrelation function for zero time delay, g$^{(2)}$($0$), will
then be below 0.5~\cite{loudon}. However, it becomes clear from the
LED based images presented above that this is not yet possible. The
presence of other molecules or background light introduces
additional photons, similar to having more emitters with a certain
detection probability of $\eta_{\rm i,det}$ in the observation spot.
The autocorrelation function for zero time delay can be expressed as
$g^{(2)}(0)=1-1/(1+\sum_{i=1}^{n} \eta_{\rm i,det} P_{\rm i})$,
where $n$ is the number of additional emitters. The signal to
background ratio (SBR) would be given by $\rm{SBR}=1/(\sum_{i=1}^{n}
\eta_{\rm i,det} P_{\rm i})$, such that the final equation reads
$g^{(2)}(0)=1/(1+\rm{SBR})$. When using higher laser intensities at
a few hundred kW/m$^2$ in a confocal configuration, we were able to
quickly acquire autocorrelation signals with values below
$g^{(2)}(0)=0.2$, thus proving the workability of the sample. In the
case of LED illumination and a signal to background ratio of 0.12,
it will not be possible to acquire an autocorrelation function below
$g^{(2)}(0) \approx 0.89$. The noise and background floor need to be
brought down by a factor of 9 for the dip in the $g^{(2)}(0)$ to
fall below 0.5 as required. This problem might be circumvented by
utilizing quartz coverslips to reduce fluorescence background and an
optimized filter set with a cut-off wavelength at 565~nm. Another
approach would be to spatially filter the emitted light more
efficiently, such as in confocal imaging or to capture the light in
a single mode fiber. In further studies we did not succeed in
detecting single molecules with LED illumination in a
\emph{confocal} configuration. The spatial filtering by single mode
fibers reduced the effective efficiency too drastically, and has
also been reported by other groups~\cite{hattori_cl_2009}.

To achieve a triggered single photon emission an optical pulse width
below the molecules T$_1$ time, {\em i.e.},~a few ns, should be
used. Such LEDs are currently available on the market (e.g.,
Picoquant, PLS-series), but the irradiance is significantly smaller
than the ones used in this paper. With the LED used here we were
able to generate short pulses down to 100ns. The bond wires'
inductance might be the limiting factor in shortening the pulse for
our device, but this was not further explored.

\section{Conclusions}

Performing single molecule spectroscopy with other illumination
sources than lasers still remains a challenging task. The current
efforts in engineering of semiconductor materials promise that
light-emitting diodes will find their way into further single
molecule studies in the future. The advantages of using LEDs are
their compact design, their robustness, their intensity stability
and their low cost, which make them a valuable tool to detect also
small amount of fluorescent samples. Other particles like quantum
dots might be even more successful candidates for LED excitation,
because of their higher absorbtion in the blue region of the
spectrum, opening the option for better spectral discrimination. It
is an interesting option to illuminate vacancy centers in diamond,
aiming for a triggered single photon source. Unfortunately usual
vacancies in diamond (NV$^-$) require a higher excitation irradiance
and the detection relies on the red-shifted emission of the phonon
wing. This makes the excitation and as well the discrimination
between fluorescence and illumination photons even harder.

\section*{Acknowledgements}
The Authors would like to thank Gleb~Maslenikov, Meng-Khoon~Tey and
Aaron~Danner for supporting the experimental tasks.

This work has been financially supported by National Research
Foundation Singapore.


\begin{thebibliography}{10}%
\makeatletter
\providecommand \@ifxundefined [1]{%
 \ifx #1\undefined \expandafter \@firstoftwo
 \else \expandafter \@secondoftwo
\fi
}%
\providecommand \@ifnum [1]{%
 \ifnum #1\expandafter \@firstoftwo
 \else \expandafter \@secondoftwo
\fi
}%
\providecommand \enquote [1]{``#1''}%
\providecommand \bibnamefont  [1]{#1}%
\providecommand \bibfnamefont [1]{#1}%
\providecommand \citenamefont [1]{#1}%
\providecommand\href[0]{\@sanitize\@href}%
\providecommand\@href[1]{\endgroup\@@startlink{#1}\endgroup\@@href}%
\providecommand\@@href[1]{#1\@@endlink}%
\providecommand \@sanitize [0]{\begingroup\catcode`\&12\catcode`\#12\relax}%
\@ifxundefined \pdfoutput {\@firstoftwo}{%
 \@ifnum{\z@=\pdfoutput}{\@firstoftwo}{\@secondoftwo}%
}{%
 \providecommand\@@startlink[1]{\leavevmode}%
 \providecommand\@@endlink[0]{}%
}{%
 \providecommand\@@startlink[1]{%
  \leavevmode
  \pdfstartlink
   attr{/Border[0 0 1 ]/H/I/C[0 1 1]}%
   user{/Subtype/Link/A<</Type/Action/S/URI/URI(#1)>>}%
  \relax
 }%
 \providecommand\@@endlink[0]{\pdfendlink}%
}%
\providecommand \url  [0]{\begingroup\@sanitize \@url }%
\providecommand \@url [1]{\endgroup\@href {#1}{\urlprefix}}%
\providecommand \urlprefix [0]{URL }%
\providecommand \Eprint[0]{\href }%
\@ifxundefined \urlstyle {%
  \providecommand \doi [1]{doi:\discretionary{}{}{}#1}%
}{%
  \providecommand \doi [0]{doi:\discretionary{}{}{}\begingroup
  \urlstyle{rm}\Url }%
}%
\providecommand \doibase [0]{http://dx.doi.org/}%
\providecommand \Doi[1]{\href{\doibase#1}}%
\providecommand \bibAnnote [3]{%
  \BibitemShut{#1}%
  \begin{quotation}\noindent
    \textsc{Key:}\ #2\\\textsc{Annotation:}\ #3%
  \end{quotation}%
}%
\providecommand \bibAnnoteFile [2]{%
  \IfFileExists{#2}{\bibAnnote {#1} {#2} {\input{#2}}}{}%
}%
\providecommand \typeout [0]{\immediate \write \m@ne }%
\providecommand \selectlanguage [0]{\@gobble}%
\providecommand \bibinfo [0]{\@secondoftwo}%
\providecommand \bibfield [0]{\@secondoftwo}%
\providecommand \translation [1]{[#1]}%
\providecommand \BibitemOpen[0]{}%
\providecommand \bibitemStop [0]{}%
\providecommand \bibitemNoStop [0]{.\EOS\space}%
\providecommand \EOS [0]{\spacefactor3000\relax}%
\providecommand \BibitemShut [1]{\csname bibitem#1\endcsname}%
\bibitem{hirschfeld01}%
  \BibitemOpen
  \bibfield{author}{%
  \bibinfo {author} {\bibfnamefont{T.}~\bibnamefont{Hirschfeld}},\ }%
  \bibfield{journal}{%
  \bibinfo {journal} {Applied Optics}\ }%
  \textbf{\bibinfo {volume} {15}},\ \bibinfo {pages} {2965} (\bibinfo {month}
  {12}\ \bibinfo {year} {1976}),\
  \url{http://www.opticsinfobase.org/abstract.cfm?URI=ao-15-12-2949}%
  \bibAnnoteFile{NoStop}{hirschfeld01}%
\bibitem{moerner01}%
  \BibitemOpen
  \bibfield{author}{%
  \bibinfo {author} {\bibfnamefont{W.~E.}\ \bibnamefont{Moerner}}\ and\
  \bibinfo {author} {\bibfnamefont{L.}~\bibnamefont{Kador}},\ }%
  \bibfield{journal}{%
  \bibinfo {journal} {Physical Review Letters}\ }%
  \textbf{\bibinfo {volume} {62}},\ \bibinfo {pages} {2535} (\bibinfo {month}
  {5}\ \bibinfo {year} {1989}),\
  \url{http://link.aps.org/abstract/PRL/v62/p2535}%
  \bibAnnoteFile{NoStop}{moerner01}%
\bibitem{orrit02}%
  \BibitemOpen
  \bibfield{author}{%
  \bibinfo {author} {\bibfnamefont{M.}~\bibnamefont{Orrit}}\ and\ \bibinfo
  {author} {\bibfnamefont{J.}~\bibnamefont{Bernard}},\ }%
  \bibfield{journal}{%
  \bibinfo {journal} {Physical Review Letters}\ }%
  \textbf{\bibinfo {volume} {65}},\ \bibinfo {pages} {2716} (\bibinfo {month}
  {11}\ \bibinfo {year} {1990}),\
  \url{http://link.aps.org/abstract/PRL/v65/p2716}%
  \bibAnnoteFile{NoStop}{orrit02}%
\bibitem{betzig07}%
  \BibitemOpen
  \bibfield{author}{%
  \bibinfo {author} {\bibfnamefont{E.}~\bibnamefont{Betzig}}\ and\ \bibinfo
  {author} {\bibfnamefont{R.~J.}\ \bibnamefont{Chichester}},\ }%
  \bibfield{journal}{%
  \bibinfo {journal} {Science}\ }%
  \textbf{\bibinfo {volume} {262}},\ \bibinfo {pages} {1422} (\bibinfo {month}
  {11}\ \bibinfo {year} {1993}),\
  \url{http://www.jstor.org/view/00368075/di002241/00p0186x/0}%
  \bibAnnoteFile{NoStop}{betzig07}%
\bibitem{Hell2003}%
  \BibitemOpen
  \bibfield{author}{%
  \bibinfo {author} {\bibfnamefont{S.~W.}\ \bibnamefont{Hell}},\ }%
  \bibfield{journal}{%
  \bibinfo {journal} {Nature Biotechnology}\ }%
  \textbf{\bibinfo {volume} {21}},\ \bibinfo {pages} {1347} (\bibinfo {year}
  {2003}),\ \url{http://dx.doi.org/10.1038/nbt895}%
  \bibAnnoteFile{NoStop}{Hell2003}%
\bibitem{unger_biotechniques_1999}%
  \BibitemOpen
  \bibfield{author}{%
  \bibinfo {author} {\bibfnamefont{M.}~\bibnamefont{Unger}}, \bibinfo {author}
  {\bibfnamefont{E.}~\bibnamefont{Kartalov}}, \bibinfo {author}
  {\bibfnamefont{C.-S.}\ \bibnamefont{Chiu}}, \bibinfo {author}
  {\bibfnamefont{H.}~\bibnamefont{Lester}},\ and\ \bibinfo {author}
  {\bibfnamefont{S.}~\bibnamefont{Quake}},\ }%
  \bibfield{journal}{%
  \bibinfo {journal} {BioTechniques}\ }%
  \textbf{\bibinfo {volume} {27}},\ \bibinfo {pages} {1008} (\bibinfo {month}
  {Nov.}\ \bibinfo {year} {1999})%
  \bibAnnoteFile{NoStop}{unger_biotechniques_1999}%
\bibitem{kuo}%
  \BibitemOpen
  \bibfield{author}{%
  \bibinfo {author} {\bibfnamefont{J.~S.}\ \bibnamefont{Kuo}}, \bibinfo
  {author} {\bibfnamefont{C.~L.}\ \bibnamefont{Kuyper}}, \bibinfo {author}
  {\bibfnamefont{P.~B.}\ \bibnamefont{Allen}}, \bibinfo {author}
  {\bibfnamefont{G.~S.}\ \bibnamefont{Fiorini}},\ and\ \bibinfo {author}
  {\bibfnamefont{D.~T.}\ \bibnamefont{Chiu}},\ }%
  \bibfield{journal}{%
  \bibinfo {journal} {Electrophoresis}\ }%
  \textbf{\bibinfo {volume} {25}},\ \bibinfo {pages} {3796} (\bibinfo {year}
  {2004}),\ \url{http://dx.doi.org/10.1002/elps.200406118}%
  \bibAnnoteFile{NoStop}{kuo}%
\bibitem{hattori_cl_2009}%
  \BibitemOpen
  \bibfield{author}{%
  \bibinfo {author} {\bibfnamefont{A.}~\bibnamefont{Hattori}}, \bibinfo
  {author} {\bibfnamefont{S.}~\bibnamefont{Habuchi}},\ and\ \bibinfo {author}
  {\bibfnamefont{M.}~\bibnamefont{Vacha}},\ }%
  \bibfield{journal}{%
  \bibinfo {journal} {Chemistry Letters}\ }%
  \textbf{\bibinfo {volume} {38}},\ \bibinfo {pages} {234} (\bibinfo {year}
  {2009}),\
  \url{http://www.jstage.jst.go.jp.libproxy1.nus.edu.sg/article/cl/38/3/38_234%
/_article}%
  \bibAnnoteFile{NoStop}{hattori_cl_2009}%
\bibitem{basche01}%
  \BibitemOpen
  \bibfield{author}{%
  \bibinfo {author} {\bibfnamefont{T.}~\bibnamefont{Basch\'e}}, \bibinfo
  {author} {\bibfnamefont{W.~E.}\ \bibnamefont{Moerner}}, \bibinfo {author}
  {\bibfnamefont{M.}~\bibnamefont{Orrit}},\ and\ \bibinfo {author}
  {\bibfnamefont{H.}~\bibnamefont{Talon}},\ }%
  \bibfield{journal}{%
  \bibinfo {journal} {Phys. Rev. Lett.}\ }%
  \textbf{\bibinfo {volume} {69}},\ \bibinfo {pages} {1516} (\bibinfo {month}
  {09}\ \bibinfo {year} {1992}),\
  \url{http://link.aps.org/abstract/PRL/v69/p1516}%
  \bibAnnoteFile{NoStop}{basche01}%
\bibitem{pfab}%
  \BibitemOpen
  \bibfield{author}{%
  \bibinfo {author} {\bibfnamefont{R.}~\bibnamefont{Pfab}}, \bibinfo {author}
  {\bibfnamefont{J.}~\bibnamefont{Zimmermann}}, \bibinfo {author}
  {\bibfnamefont{C.}~\bibnamefont{Hettich}}, \bibinfo {author}
  {\bibfnamefont{I.}~\bibnamefont{Gerhardt}}, \bibinfo {author}
  {\bibfnamefont{A.}~\bibnamefont{Renn}},\ and\ \bibinfo {author}
  {\bibfnamefont{V.}~\bibnamefont{Sandoghdar}},\ }%
  \bibfield{journal}{%
  \bibinfo {journal} {Chemical Physics Letters}\ }%
  \textbf{\bibinfo {volume} {387}},\ \bibinfo {pages} {490} (\bibinfo {year}
  {2004}),\ \url{http://dx.doi.org/10.1016/j.cplett.2004.02.040}%
  \bibAnnoteFile{NoStop}{pfab}%
\bibitem{kummer}%
  \BibitemOpen
  \bibfield{author}{%
  \bibinfo {author} {\bibfnamefont{S.}~\bibnamefont{Kummer}}, \bibinfo {author}
  {\bibfnamefont{F.}~\bibnamefont{Kulzer}}, \bibinfo {author}
  {\bibfnamefont{R.}~\bibnamefont{Kettner}}, \bibinfo {author}
  {\bibfnamefont{T.}~\bibnamefont{Basch\'e}}, \bibinfo {author}
  {\bibfnamefont{C.}~\bibnamefont{Tietz}}, \bibinfo {author}
  {\bibfnamefont{C.}~\bibnamefont{Glowatz}},\ and\ \bibinfo {author}
  {\bibfnamefont{C.}~\bibnamefont{Kryschi}},\ }%
  \bibfield{journal}{%
  \Doi{10.1063/1.475107}{\bibinfo {journal} {The Journal of Chemical Physics}}\
  }%
  \textbf{\bibinfo {volume} {107}},\ \bibinfo {pages} {7673} (\bibinfo {year}
  {1997}),\ \url{http://link.aip.org/link/?JCP/107/7673/1}%
  \bibAnnoteFile{NoStop}{kummer}%
\bibitem{kulzer_chemphys_1999}%
  \BibitemOpen
  \bibfield{author}{%
  \bibinfo {author} {\bibfnamefont{F.}~\bibnamefont{Kulzer}}, \bibinfo {author}
  {\bibfnamefont{F.}~\bibnamefont{Koberling}}, \bibinfo {author}
  {\bibfnamefont{T.}~\bibnamefont{Christ}}, \bibinfo {author}
  {\bibfnamefont{A.}~\bibnamefont{Mews}},\ and\ \bibinfo {author}
  {\bibfnamefont{T.}~\bibnamefont{Basch\'e}},\ }%
  \bibfield{journal}{%
  \bibinfo {journal} {Chemical Physics}\ }%
  \textbf{\bibinfo {volume} {247}},\ \bibinfo {pages} {23} (\bibinfo {month}
  {August}\ \bibinfo {year} {1999}),\
  \url{http://dx.doi.org/10.1016/S0301-0104(99)00100-7}%
  \bibAnnoteFile{NoStop}{kulzer_chemphys_1999}%
\bibitem{thompson_biophysjourn_2002}%
  \BibitemOpen
  \bibfield{author}{%
  \bibinfo {author} {\bibfnamefont{R.~E.}\ \bibnamefont{Thompson}}, \bibinfo
  {author} {\bibfnamefont{D.~R.}\ \bibnamefont{Larson}},\ and\ \bibinfo
  {author} {\bibfnamefont{W.~W.}\ \bibnamefont{Webb}},\ }%
  \bibfield{journal}{%
  \bibinfo {journal} {Biophysical Journal}\ }%
  \textbf{\bibinfo {volume} {82}},\ \bibinfo {pages} {27752783} (\bibinfo
  {year} {2002}),\ \url{http://dx.doi.org/10.1016/S0006-3495(02)75618-X}%
  \bibAnnoteFile{NoStop}{thompson_biophysjourn_2002}%
\bibitem{peter_pss_2009}%
  \BibitemOpen
  \bibfield{author}{%
  \bibinfo {author} {\bibfnamefont{M.}~\bibnamefont{Peter}}, \bibinfo {author}
  {\bibfnamefont{A.}~\bibnamefont{Laubsch}}, \bibinfo {author}
  {\bibfnamefont{W.}~\bibnamefont{Bergbauer}}, \bibinfo {author}
  {\bibfnamefont{T.}~\bibnamefont{Meyer}}, \bibinfo {author}
  {\bibfnamefont{M.}~\bibnamefont{Sabathil}}, \bibinfo {author}
  {\bibfnamefont{J.}~\bibnamefont{Baur}},\ and\ \bibinfo {author}
  {\bibfnamefont{B.}~\bibnamefont{Hahn}},\ }%
  \bibfield{journal}{%
  \bibinfo {journal} {phys. stat. sol. (a)}\ }%
  \textbf{\bibinfo {volume} {206}},\ \bibinfo {pages} {1125} (\bibinfo {year}
  {2009}),\ ISSN \bibinfo {issn} {1862-6319},\
  \url{http://dx.doi.org/10.1002/pssa.200880926}%
  \bibAnnoteFile{NoStop}{peter_pss_2009}%
\bibitem{kukura-2008}%
  \BibitemOpen
  \bibfield{author}{%
  \bibinfo {author} {\bibfnamefont{P.}~\bibnamefont{Kukura}}, \bibinfo {author}
  {\bibfnamefont{M.}~\bibnamefont{Celebrano}}, \bibinfo {author}
  {\bibfnamefont{A.}~\bibnamefont{Renn}},\ and\ \bibinfo {author}
  {\bibfnamefont{V.}~\bibnamefont{Sandoghdar}},\ }%
  \bibfield{journal}{%
  \bibinfo {journal} {Nano Letters}\ }%
  \textbf{\bibinfo {volume} {9}},\ \bibinfo {pages} {926929} (\bibinfo {year}
  {2008}),\ \url{http://arxiv.org/abs/0802.1206}%
  \bibAnnoteFile{NoStop}{kukura-2008}%
\bibitem{moerner04}%
  \BibitemOpen
  \bibfield{author}{%
  \bibinfo {author} {\bibfnamefont{W.~E.}\ \bibnamefont{Moerner}}, \bibinfo
  {author} {\bibfnamefont{T.}~\bibnamefont{Plakhotnik}}, \bibinfo {author}
  {\bibfnamefont{T.}~\bibnamefont{Irngatinger}}, \bibinfo {author}
  {\bibfnamefont{M.}~\bibnamefont{Croci}}, \bibinfo {author}
  {\bibfnamefont{V.}~\bibnamefont{Palm}},\ and\ \bibinfo {author}
  {\bibfnamefont{U.~P.}\ \bibnamefont{Wild}},\ }%
  \bibfield{journal}{%
  \bibinfo {journal} {J. Phys. chem.}\ }%
  \textbf{\bibinfo {volume} {98}},\ \bibinfo {pages} {7382} (\bibinfo {year}
  {1994}),\ \url{http://dx.doi.org/10.1021/j100081a025}%
  \bibAnnoteFile{NoStop}{moerner04}%
\bibitem{Bushberg2001}%
  \BibitemOpen
  \bibfield{author}{%
  \bibinfo {author} {\bibfnamefont{J.~T.}\ \bibnamefont{Bushberg}}, \bibinfo
  {author} {\bibfnamefont{J.~A.}\ \bibnamefont{Seibert}}, \bibinfo {author}
  {\bibfnamefont{E.~M.~L.}\ \bibnamefont{Jr}},\ and\ \bibinfo {author}
  {\bibfnamefont{J.~M.}\ \bibnamefont{Boone}},\ }%
  \emph{\bibinfo {title} {The Essential Physics of Medical Imaging}},\ \bibinfo
  {edition} {2nd}\ ed.\ (\bibinfo {publisher} {Lippincott Williams \&
  Wilkins},\ \bibinfo {year} {2001})\ ISBN \bibinfo {isbn} {0683301187}%
  \bibAnnoteFile{NoStop}{Bushberg2001}%
\bibitem{loudon}%
  \BibitemOpen
  \bibfield{author}{%
  \bibinfo {author} {\bibfnamefont{R.}~\bibnamefont{Loudon}},\ }%
  \emph{\bibinfo {title} {The Quantum Theory of Light}}\ (\bibinfo {publisher}
  {Oxford University Press},\ \bibinfo {year} {2000})%
  \bibAnnoteFile{NoStop}{loudon}%
\end{thebibliography}
\end{document}